\documentclass[preprint]{elsarticle}
\usepackage{amsmath,amssymb}
\usepackage{slashed}
\usepackage{graphicx}%
\usepackage{subfigure}
\usepackage{mathptmx}
\usepackage{bm}
\newcommand{\beq}{\begin{equation}}
\newcommand{\eeq}{\end{equation}}

\newcommand{\be}{\begin{eqnarray}}
\newcommand{\ee}{\end{eqnarray}}
\long\def\hidestart#1\hideend{}
\begin{document}

\title{Chiral Anomaly in Lattice QCD with Twisted Mass Wilson Fermion}

\author{Asit K. De\corref{cor1}}
\ead{asitk.de@saha.ac.in}

\author{A. Harindranath}
\ead{a.harindranath@saha.ac.in}

\author{Santanu Mondal}
\ead{santanu.mondal@saha.ac.in}

\address{Theory Division, Saha Institute of Nuclear Physics \\
 1/AF Bidhan Nagar, Kolkata 700064, India}
\cortext[cor1]{Corresponding author}
\date{October 21, 2009}
\begin{abstract}

The flavour singlet axial Ward identity with Osterwalder-Seiler twisted mass
Wilson fermion action is studied on a finite lattice, with finite fermion mass
and the Wilson parameter $r$ up to 1. Approach to the infinite volume chiral
limit and emergence of the anomaly is significantly better than that obtained with ${\cal
O}(a)$ and ${\cal O}(a^2)$ improved fermion actions. We have shown
explicitly that up to ${\cal O}(g^2)$, parity violating terms cancel in the
Ward identity even at finite volume and finite lattice spacing.

\end{abstract}

\maketitle
\section{Introduction}\label{intro}
It is well known that the naive Wilson fermion \cite{wilson1} reproduces 
the chiral 
anomaly  in lattice QCD in the continuum limit which was demonstrated 
analytically in 
weak coupling lattice perturbation theory by Karsten and Smit 
\cite{karsten-smit} who also showed that the result is independent of 
the Wilson parameter $r$. Subsequently, Kerler \cite{kerler1} starting 
from the flavor singlet axial Ward Identity also 
showed the emergence of the 
anomaly term  in the continuum limit for small values of $r$.

The Wilson fermions have a dimension five chiral symmetry breaking term 
which leads to complications in lattice simulations, such as additive quark mass
renormalization, axial current renormalization and nontrivial mixing between 
operators. Even though the ${\cal O}(a)$ lattice artifacts associated with
Wilson fermions go to zero in the continuum limit, the technical difficulties
are significant in actual simulations which are performed in a finite volume
with lattice spacing and quark mass both being non-zero. 

Currently, twisted mass Wilson QCD (tm WQCD) 
(for reviews, see, for example, \cite{frezzotti,sint,shindler}) 
has become a popular way
of  simulating lattice QCD. In addition to the absence of 
unphysical zero modes of the
Wilson-Dirac operator, the ability to overcome some of the lattice
renormalization problems and the automatic ${\cal O}(a)$ improvement
are its obvious advantages.

In the early days of lattice QCD, Osterwalder and Seiler 
\cite{osterwalder-seiler} introduced a
chirally twisted Wilson term (OStm WQCD) to avoid doublers. Compared
to OStm WQCD, tm WQCD has a nontrivial flavour dependence which leads
to flavour and parity breaking in the lattice theory which is expected
to disappear in the continuum limit. On the other hand, parity
violating effects of OStm WQCD may survive the continuum limit
non-perturbatively.

In this work we explicitly show that
the parity violating terms are absent up to ${\cal O}(g^2)$ in the
flavour singlet axial Ward Identity on the {\em finite} lattice. 
We study the dependence of the
chiral anomaly with OStm WQCD on the Wilson parameter $r$ as a
function of the
lattice quark mass $am$  and show the independence as one approaches the
chiral limit. We also study the volume dependence of
the result as one approaches the chiral limit.

The chiral anomaly on the lattice with finite 
volume and non-zero quark mass actually provides an excellent laboratory to compare
the OStm WQCD with unimproved as well as ${\cal O}(a)$
and ${\cal O}(a^2)$ 
improved actions with Wilson fermions. Our detailed
investigation provides a quantitative measure of  the effectiveness of the
twisted mass Wilson fermions in further
reducing finite lattice spacing artifacts compared to  ${\cal O}(a)$ and
${\cal O}(a^2)$  \cite{hamber-wu, wetzel, eguchi-kawamoto} 
improved Wilson fermions.    

Two comments, however, are in order: (i) The weak coupling perturbative analysis
is not sensitive to
some lattice artifacts that may be present in numerical simulations
(for example, see \cite{ETMC09}), (ii) the positivity property of 
the quark determinant is not
satisfied for the OStm WQCD action \cite{frezzotti_jhep} and hence
numerical simulation with dynamical quarks with this action is not feasible.
Numerical simulation is nevertheless possible with a mixed action (the OStm WQCD action for
the valence quarks and the tm WQCD action 
for the sea quarks).   
\section{Osterwalder-Seiler Chirally Twisted Wilson Term}
The standard Wilson fermion action
\begin{eqnarray}
 S_F[\psi,{\overline\psi},U] =a^4~\sum_{x,y}{\overline \psi}_{x}
M_{xy}\psi_{y} = 
  a^4~\sum_{x,y}{\overline \psi}_{x} 
\left [ \gamma_\mu D_\mu + W +m \right ]_{xy}\psi_{y}
\end{eqnarray}
where
\begin{eqnarray}
[D_\mu]_{xy} &=&  \frac{1}{2a}~\left [U_{x,\mu }~\delta_{x+\mu,y}  - 
U^\dagger_{x-\mu, \mu}~\delta_{x-\mu,y} \right ]~,\nonumber \\
W_{xy} &=&  \frac{r}{2a}~\sum_\mu \left [ 2 \delta_{x,y} - 
U_{x,\mu }~\delta_{x+\mu,y}  - 
U^\dagger_{x-\mu, \mu}~\delta_{x-\mu,y} \right ]~,
\end{eqnarray}
with the variations $
\psi_x \rightarrow \psi'_x = \left [ 1 - i \gamma_5 \alpha_x \right ]
\psi_x, \, \, \, 
{\overline \psi}_x \rightarrow {\overline \psi}'_x = {\overline \psi}_x
\left [ 1 - i \gamma_5 \alpha_x \right ] $
lead to the flavor singlet axial Ward Identity 
\begin{eqnarray}
\langle {\Delta}^{b}_{\mu} J_{5 \mu}(x) \rangle = 2m \langle 
{\overline  \psi}_x \gamma_5 \psi_x \rangle + \langle \chi_x\rangle 
\end{eqnarray}
where $\langle {\cal O} \rangle $ denotes the functional
average of ${\cal O}$. Explanation of other terms are as follows:
\begin{eqnarray}
{\rm The~ backward \, \, derivative,}~~~
\Delta^b_\mu f(x) &=& \frac{1}{a} \left  [ f(x) - f(x -\mu) \right ]~,
 \nonumber  \\ 
J_{5 \mu}(x) &=& \frac{1}{2} \left [ 
{\overline \psi}_x \gamma_\mu \gamma_5 U_{x,\mu} \psi_{x+\mu} 
+ {\overline \psi}_{x+\mu} \gamma_\mu \gamma_5 U^\dagger_{x \mu}
\psi_x \right ]  \nonumber \\
~~{\rm and}~~ \langle \chi_x \rangle &=& -{\rm Trace} [\gamma_5 (GW+WG)]~.
\end{eqnarray}
The Green's function 
\begin{eqnarray}
G(x,y) = \langle x \mid \frac{1}{[\gamma_\mu D_\mu +W+m]} \mid y \rangle ~.
\end{eqnarray}
Following the method of Kerler \cite{kerler1}, to ${\cal O}(g^2)$, one
arrives at
\begin{eqnarray}
\langle \chi_x \rangle &=& 2 ~ g^2 ~ \epsilon_{\mu \nu \rho \lambda} ~ 
F_{\mu \nu}(x) F_{\rho \lambda}(x)~ \frac{1}{(2 \pi)^4}\sum_p 
{\rm cos}(p_{\mu}a){\cos}(p_{\nu}a) {\cos}(p_{\rho}a)~ \nonumber \\ 
& {\hspace{.2in}}\times & W_0(p) \Big [ {\cos}(p_\lambda a) [ m + W_0(p)] 
                             - ~ 4 r {\sin}^2 (p_\lambda a) \Big ]
({\cal G}_0(p))^3 , \label{caw1} \\
&=& - \frac{g^2}{16 \pi^2} ~ \epsilon_{\mu \nu \rho \lambda}~ 
F_{\mu \nu}(x) F_{\rho \lambda}(x) I(am,r,L) \label{caw2}
\end{eqnarray}
Here 
\begin{eqnarray}
W_{0}(p) &= &\frac{r}{a} \sum_\mu [ (1 - {\rm cos}(a p_\mu) ) ]~,
\nonumber \\
{\cal G}_0(p) &=& \left (   
\frac{1}{a^2} \sum_\mu {\sin}^2 (a p_\mu ) + (m + \frac{r}{a} \sum_\mu 
[ 1 - {\cos}(a p_\mu )])^2
\right )^{-1}.
\end{eqnarray}
Explicitly, $ \sum_p = (\frac{1}{L})^4 \sum_{n_1,n_2,n_3,n_4}$ 
where $n_1,n_2,n_3,n_4=0,1,2,3, \cdots$. In the infinite volume chiral limit, 
$I \rightarrow 1 $. In all our plots it is 
the function $I(am,r,L)$ which we have plotted.

Next we consider the chirally twisted Wilson term introduced by 
Osterwalder and Seiler \cite{osterwalder-seiler} 
which amounts to the replacement 
$W \rightarrow R=-i\gamma_5 W$ \cite{kerler1}. (Note that Seiler and
Stamatescu \cite{seiler-stamatescu} introduced the generalization $-i
\gamma_5 \rightarrow {\rm exp}(-i
\theta \gamma_5)$ so that the chiral angle $\theta$ can be directly
related to the theta vacua of QCD. This generalization is {\em not}
considered in this work.)

The OStm WQCD leads to the Flavor Singlet Axial Ward Identity
\begin{eqnarray}
\langle \Delta^b_\mu J_{\mu 5}(x)\rangle &=& 
 -2m ~{\rm Trace} ~ \gamma_5 G^{os} - \langle \chi^{os}(x)\rangle \nonumber \\ 
&=& -2m ~ {\rm Trace} ~ \gamma_5 G^{os}~ -~
{\rm Trace}~ \gamma_5 (G^{os}R+RG^{os}) 
\end{eqnarray} 
with 
\begin{eqnarray}
G^{os} = \frac{1}{{\slashed D} +m -i \gamma_5W} 
      = ({\slashed D} - m - i \gamma_5 W) \frac{1}{({\cal G}^{os})^{-1} -V}
\end{eqnarray}
where
\begin{eqnarray}
({\cal G}^{os})^{-1} = D^2 - m^2 - W^2 ~ {\rm and}~ V= 
V_1+V_2^{os} \nonumber \\
{\rm with}~ V_1=\frac{i}{2}\sigma_{\mu \nu}
[D_\mu,D_\nu] ~~{\rm and}~~ V_2^{os}=- i \gamma_5 [{\slashed D},W]~.
\end{eqnarray}
Evaluation of $\langle \chi^{os}(x)\rangle$ following a similar
method as above now produces some 
parity violating terms which are not of the form $F{\tilde F}$ and therefore 
do not contribute to the anomaly, however, would contribute to the axial
Ward identity.
Up to ${\cal O} (g^2)$ these parity violating terms arising out of $\langle
\chi^{os}(x)\rangle$ are as follows,
\begin{eqnarray}
{\rm {\cal O}} ~(g^0):\hspace{1cm}&& -2im ~{\rm Trace} ~W{\cal G}^{os}
\nonumber \\
{\rm {\cal O}} ~(g^1):\hspace{1cm} 
&&{\rm vanishes ~~because ~~of ~~Dirac ~~ Trace} \nonumber \\
{\rm {\cal O}} ~(g^2):\hspace{1cm} 
&& -2im~ {\rm Trace}~W{\cal G}^{os} V_1{\cal G}^{os}V_1
{\cal G}^{os} - 2im~ {\rm Trace}~W{\cal G}^{os} V_2^{os}{\cal G}^{os}V_2^{os}
{\cal G}^{os} ~.\nonumber \\ \label{pvc}
\end{eqnarray}
 
Contribution to the anomaly from $\langle \chi^{os}(x)\rangle$ can be
calculated by making
the following changes from the standard Wilson case:
\be
{\rm Trace}~\gamma_5 (W+m){\cal G}V_1{\cal G}V_1 {\cal G}W 
\rightarrow {\rm Trace}~\gamma_5 W{\cal G}^{os}V_1{\cal G}^{os}V_1 
{\cal G}^{os}W   
\ee 
and
\be
{\rm Trace} ~\gamma_5 {\slashed D} {\cal G}V_1{\cal G}[{\slashed D},W]
{\cal G}W \rightarrow  {\rm Trace} ~\gamma_5 {\slashed D} {\cal G}^{os}
V_1{\cal G}^{os}[{\slashed D},W] {\cal G}^{os}W~.
\ee
Thus compared to the expression in the case of 
standard Wilson term, the expression for anomaly in the case 
of Osterwalder-Seiler  
Wilson term has the following features: (i) absence of fermion mass $m$ 
in the numerator and (ii) absence of mixing between
mass term and Wilson term ($Wm$) in the denominator.
Explicitly, in place of Eq. (\ref{caw1}) we find 
 the contribution  
to $F{\tilde F}$ from $\langle \chi^{os}(x)\rangle$, in the case of the Osterwalder-Seiler twisted Wilson term 
\begin{eqnarray}
\langle \chi^{os}_x \rangle &=& 2 ~ g^2 ~ \epsilon_{\mu \nu \rho \lambda} ~ 
F_{\mu \nu}(x) F_{\rho \lambda}(x)~ \sum_p 
{\rm cos}(p_{\mu}a){\cos}(p_{\nu}a) {\cos}(p_{\rho}a)~ \nonumber \\ 
& \times & W_0(p) \Big [ {\cos}(p_\lambda a) W_0(p) 
                             - ~ 4 r {\sin}^2 (p_\lambda a) \Big ]
({\cal G}^{os}_0(p))^3~. \label{caosw1} 
\end{eqnarray} 
In the following we refer to the terms proportional to
${\cos}(p_\lambda a) $ and ${\sin}^2 (p_\lambda a)$ as DD and DW terms
respectively since they correspond to contributions from $V_1 V_1$ and
$V_1 V_2^{os}$ terms in the trace.

In the Axial Ward Identity, the mass term can be written as
\be
-2m ~{\rm Trace}~\gamma_5 G^{os} &=& -2m ~{\rm Trace}~ \gamma_5 
({\slashed D}- m - i\gamma_5 W)({\cal G}^{os} +{\cal G}^{os} V {\cal
  G}^{os}\nonumber \\ 
&{\hspace{.5in}}&+{\cal G}^{os}V{\cal G}^{os}V{\cal G}^{os}+ \cdots ).
\ee

Up to ${\cal O} (g^2)$, the parity violating 
contributions from the mass term are
\begin{eqnarray}
{\rm {\cal O}} ~(g^0):\hspace{1cm}&&+2im ~{\rm Trace}~ W{\cal G}^{os}
\nonumber \\
{\rm {\cal O}} ~(g^1):\hspace{1cm} 
&&- 2im ~{\rm Trace} ~{\slashed D} {\cal G}^{os} 
[{\slashed D},W] {\cal G}^{os} \nonumber \\
{\rm {\cal O}} ~(g^2):\hspace{1cm} 
&&+ 2im~ {\rm Trace}~W{\cal G}^{os} V_1{\cal G}^{os}V_1
{\cal G}^{os} + 2im~ {\rm Trace}~W{\cal G}^{os} V_2^{os}{\cal G}^{os}V_2^{os}
{\cal G}^{os} ~.\nonumber \\ \label{pvm} 
\end{eqnarray} 
Comparing Eqs. (\ref{pvc}) and (\ref{pvm}), we immediately see that 
${\cal O}(g^0$) and ${\cal O} (g^2)$ parity violating 
terms cancel between mass term and $\langle \chi^{os}(x)\rangle$.
  
The ${\cal O} (g^1$) term in Eq. (\ref{pvm})
\be
\sum ({\slashed D}_0)_{xx_{1}} ({\cal G}_{0})_{x_{1}x_{2}} 
([{\slashed D},W]])_{x_{2}x_{3}}({\cal G}_{0})_{x_{3}x}  \nonumber  \\
\Longrightarrow \sin(ap_\mu) F_{\mu \rho} \Big [ (\cos(a p_\mu)+i \sin(ap_\mu))
(\cos(a p_\rho)+i \sin(ap_\rho)) \nonumber \\ 
-(\cos(a p_\mu)-i \sin(ap_\mu))
(\cos(a p_\rho)-i \sin(ap_\rho)) \nonumber \\
-(\cos(a p_\mu)+i \sin(ap_\mu))
(\cos(a p_\rho)-i \sin(ap_\rho)) \nonumber \\
+(\cos(a p_\mu)-i \sin(ap_\mu))
(\cos(a p_\rho)+i \sin(ap_\rho) \Big ] \nonumber \\
\Longrightarrow 0, ~{\rm on ~ summation}.
\ee
Thus we explicitly verify, in the case of Osterwalder-Seiler twisted Wilson
term, the cancellation of parity violating terms
in the flavour singlet Axial Ward Identity up to ${\cal O}(g^2$).
\section{Comparison with unimproved, ${\cal O}(a)$ and ${\cal O}(a^2)$
improved Wilson fermions}  
\begin{figure}
\includegraphics[width=4in,clip]{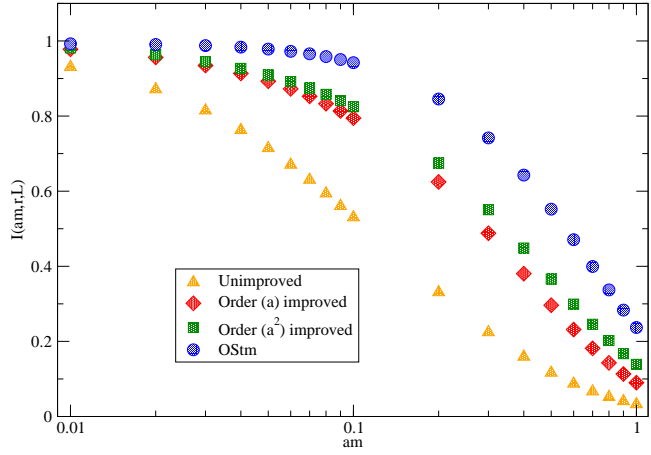}
\caption{The function $I(am,r,L)$for $r=1$ for unimproved,  
${\cal O} (a)$ improved,
${\cal O} (a^2)$ improved, and OStm Wilson fermions for the 
range of $am$ 
between 0.01 and 1.0 at $L=40$.}
\label{compua1a2ostm}
\end{figure}

Consider the following next to nearest neighbour interaction term added 
to modify the standard Wilson term \cite{hamber-wu}:
\begin{eqnarray}
\Delta S^{I} &=& a^4 \sum_{x,\mu} \Bigg \{
\frac{r}{8a} \Big [ 
{\overline \psi}(x)U_{x,\mu} U_{x+\mu,\mu} \psi(x+2 \mu) \nonumber \\
&{\hspace{.2in}}
+& {\overline \psi}(x+2 \mu) U^\dagger_{x+\mu,\mu} U^\dagger_{x,\mu} \psi(x) -2
{\overline \psi}(x) \psi(x)
\Big ]
\Bigg \}~ \nonumber \\
~~& =& a^4 \sum_x {\overline \psi}(x) W^{I}\psi(x)~. 
\end{eqnarray} 
The coefficient $\frac{r}{8a}$ is chosen so as to cancel ${\cal O}(a$) 
contributions to the tree level fermion propagator and the 
fermion-gluon vertex.
Now the total fermion action is $S=S_F+\Delta S^{I}$.
From the corresponding flavour singlet axial Ward identity, we find
the total contribution to the axial anomaly
\begin{eqnarray}
 \langle \chi_x^{I} \rangle & = &
2 \epsilon_{\mu \nu \rho \lambda}~g^2~F_{\mu \nu}(x) F_{\rho \lambda}(x) ~
\sum_p \cos(ap_\mu) \cos(ap_\rho)\cos(ap_\lambda) ~
[{\cal G}^I_0(p)]^3\nonumber \\
&\hspace{.4in}&\Big [\cos(ap_\nu)[m+ W_0(p)+W^I_0(p)] 
\nonumber \\
&\hspace{.8in}& - 4 r \sin(ap_\nu)
(\sin(ap_\nu) - \frac{1}{2} \sin(2 a p_\nu))\Big ] [W_0(p)+ W^I_0(p)]
 ~.
\end{eqnarray}
Here
\begin{eqnarray}
W_{0}(p)+W^I_0(p) = \Big [   
\sum_\mu \left [\frac{r}{a} (1 - {\rm cos}(a p_\mu) )+ 
\frac{r}{4a} (-1 + {\rm cos}(2 a p_\mu) )
\right ] \Big ],~\nonumber \\
{\cal G}^I_0(p)  = \left (   
\frac{1}{a^2} \sum_\mu {\rm sin}^2 (a p_\mu ) + (m +  \sum_\mu 
\frac{r}{a}[ 1 - {\rm cos} (ap_\mu )] +
\frac{r}{4a}[ -1 + {\rm cos}(2a p_\mu )] )^2
\right )^{-1}. 
\end{eqnarray}

Next consider the modification of the kinetic term with the next to
nearest neighbour interaction \cite{hamber-wu,wetzel, eguchi-kawamoto}:   
\begin{eqnarray}
\Delta S^{II} &=& a^4 \sum_{x,\mu} \Bigg \{
- ~\frac{1}{16a} \Big [ 
{\overline \psi}(x)\gamma_\mu U_{x,\mu} U_{x+\mu,\mu} \psi(x+2 \mu) \nonumber \\
&{\hspace{.3in}}-& {\overline \psi}(x+2 \mu) \gamma_\mu U^\dagger_{x+\mu,\mu} 
U^\dagger_{x,\mu} \psi(x) 
\Big ].
\Bigg \}~ \nonumber \\
~~& =& a^4 \sum_{x \mu} {\overline \psi}(x) \gamma_\mu D^{I}_\mu\psi(x)~ 
\end{eqnarray}
so that the total fermion action $S=S_F + \Delta  S^{I}+ \Delta
S^{II}$ is ${\cal O}(a^2$) improved and leads to the total contribution to the chiral anomaly
\begin{eqnarray}
 \langle \chi_x^{II} \rangle  = 
2 \epsilon_{\mu \nu \rho \lambda}~g^2~F_{\mu \nu}(x) F_{\rho \lambda}(x) ~
\sum_p \Pi_{\lambda=1}^4(\cos(ap_\lambda) - \frac{1}{4} \cos(2ap_\lambda)) ~
[{\cal G}^{II}_0(p)]^3\nonumber \\
\hspace{.5in} \times \Big [ 3 \frac{r}{a}-(W_0(p)+W^{I}_0(p)) \Big ] 
\times 
\Bigg [ \frac{3}{4}m + 3 \frac{r}{a} - \frac{r}{a} \sum_\mu
\nonumber \\
\hspace{.2in}  
\frac{(\cos(ap_\mu) - \frac{1}{4} \cos(2ap_\mu))^2 
+(- {ia}(D+D^I)_0)_\mu(p)(\sin(ap_\mu) - \frac{1}{2} \sin(2 a p_\mu)) }
{(\cos(ap_\mu) - \frac{1}{4} \cos(2ap_\mu))} \Bigg ]~, \nonumber \\
\end{eqnarray}
with ${(D+D^I)_0}_\lambda(p) = \frac{i}{a}
 [ \sin(p_\lambda a) - \frac{1}{8}\sin(2 p_\lambda a)]  $ and 
\begin{eqnarray}
{\cal G}^{II}_0(p) = - &\left [  
\frac{1}{a^2} \sum_\mu  \{((D+D^I)_0)_\mu)(p)\}^2 + 
\left\{\frac{3}{4}m +  W_0(p)+W_0^{I}(p) \right\}^2
\right ]^{-1}
\end{eqnarray}

Fig. \ref{compua1a2ostm} compares the function $I(am,r,L)$ 
for $r=1.0$ for 
unimproved, ${\cal O}(a)$ improved, ${\cal O}(a^2)$ improved and OStm
Wilson fermions for the range of $am$ 
between 0.01 and 1.0 at $L=40$. Note that for clarity, $am$ is plotted in the 
logarithmic scale. The figure shows that in the range of $am$ studied, 
${\cal O}(a^2)$ corrections are relatively small compared to the 
${\cal O}(a)$ corrections and the approach to the chiral limit is the fastest
and the flattest
for the OStm Wilson fermions.
\section{Numerical Results for Osterwalder-Seiler twisted mass Wilson Fermion}
In this section we present numerical results for the function $I(am,r,L)$  
for Osterwalder-Seiler twisted mass Wilson fermions. Since our main
concern is the  approach to the infinite volume chiral limit in the
continuum, we study in detail
the quark mass ($am$), the Wilson parameter ($r$) and the 
finite volume ($L$) dependence. Since a positive
semi-definite transfer matrix is guaranteed only for $r \le 1$, we restrict 
our study to the range $0 \le r \le 1 $.

\subsection{Quark mass dependence}
\begin{figure}
\includegraphics[width=4in,clip]{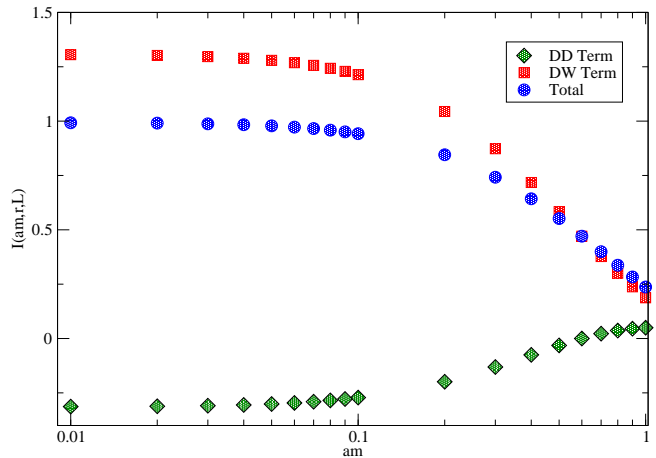}
\caption{The function $I(am,r,L)$ for $r=1$ for OStm Wilson fermions 
for the range of $am$ 
between 0.01 and 1.0 for L=40.}
\label{ostm1}
\end{figure}

In Fig. \ref{ostm1} we show the function $I(am,r,L)$ 
for OStm Wilson fermion for the range of $am$ 
between 0.1 and 1.0 at $L=40$. We find that for most values of $am$,
DW term overshoots the answer, which is compensated by the DD
term. Cutoff effects are present in the mass range $am > 0.1$, but
the result is very close to the continuum chiral answer for $am <
0.1$. 

\subsection{Wilson parameter $r$ dependence}
\begin{figure}
\subfigure{
\includegraphics[width=2.8in,clip]{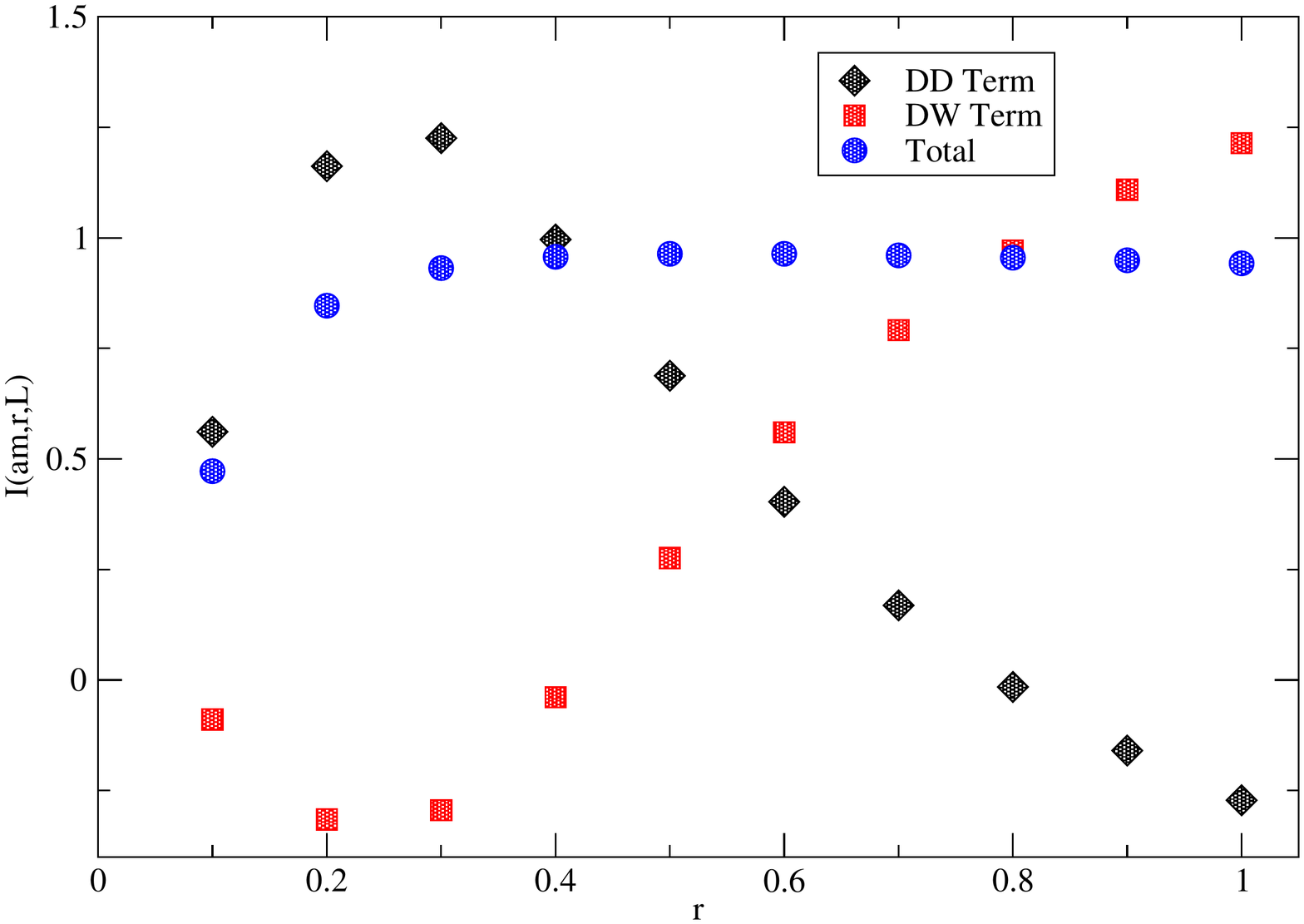}}
\subfigure{
\includegraphics[width=2.8in,clip]{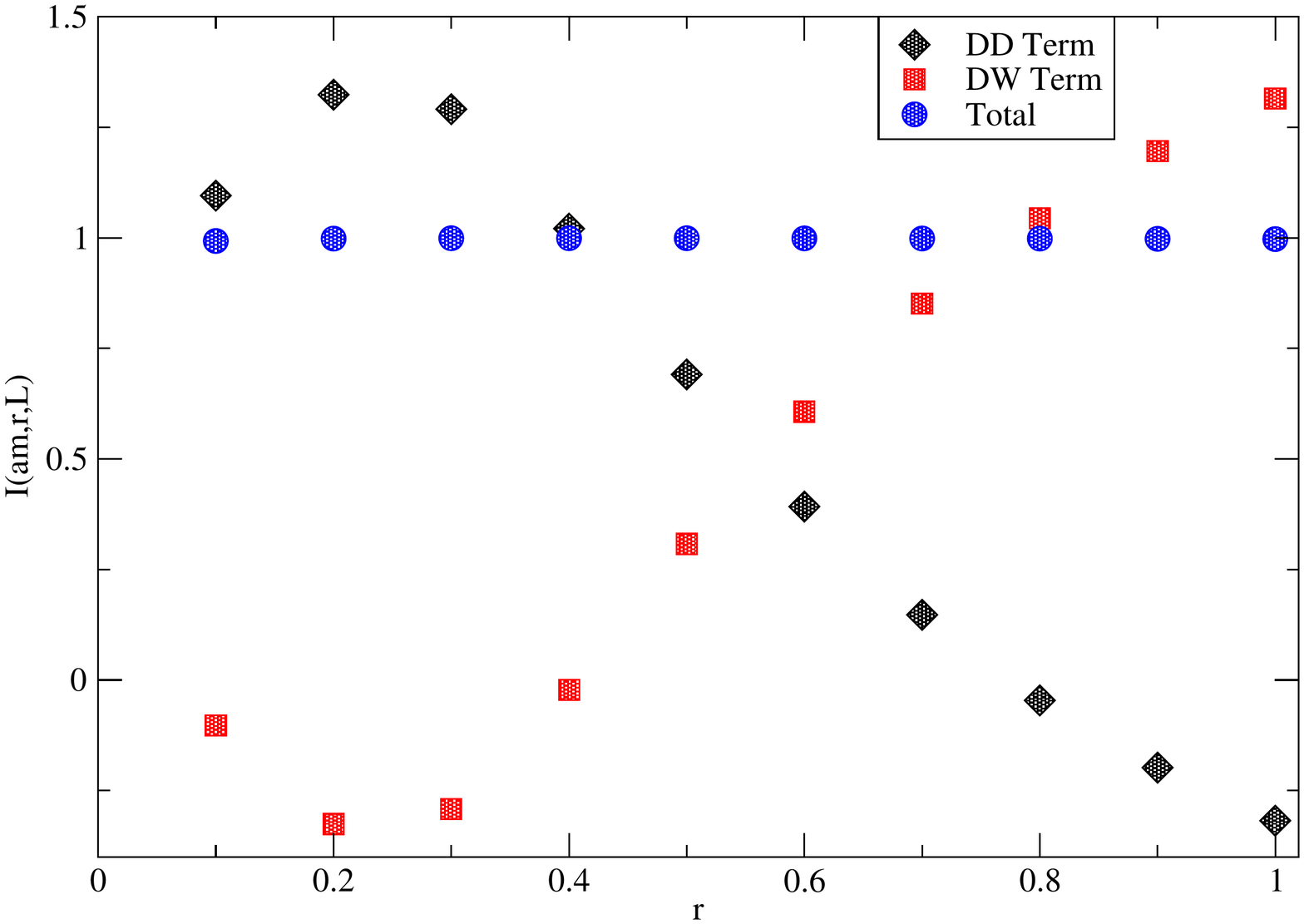}}
\caption{The function $I(am,r,L)$ for OStm Wilson fermions 
for the range of $r$ 
between 0.1 and 1.0 for  $am$ 0.01 (right) and 0.1 (left) at $L=80$. 
The contributions from DD and DW terms
are also shown separately.}
\label{ostm-rdep}
\end{figure}

For unimproved, ${\cal O}(a)$ and ${\cal O}(a^2)$ improved Wilson
fermions, it has been demonstrated that in the chiral limit, the 
function $I(am,r,L)$ is independent of the Wilson parameter $r$. For
the OStm
Wilson fermions, it is also interesting to 
check the $r$ dependence for non-zero $am$. In Fig. \ref{ostm-rdep} 
we show the $r$ dependence of terms with integrands proportional to 
$\cos(p_{\lambda}a)$ (DD) and $\sin^2(p_{\lambda}a) $ (DW) in Eq.(\ref{caosw1}) 
 and the sum 
of the  two contributions to the function $I(am,r,L)$ (DD and DW) 
for $am=.01$ (right) and $am=.1$ (left) for 
OStm  Wilson fermions at $L=80$. We note that the first term (DD) is 
dominant at small $r$ and the second term (DW) is dominant at large $r$. 
Even though the two individual contributions have very
strong $r$ dependence, the sum is seen to be independent of $r$ to a good 
numerical accuracy for $am=.01$.   
\subsection{Finite volume dependence}

\begin{figure}
\includegraphics[width=4in,clip]{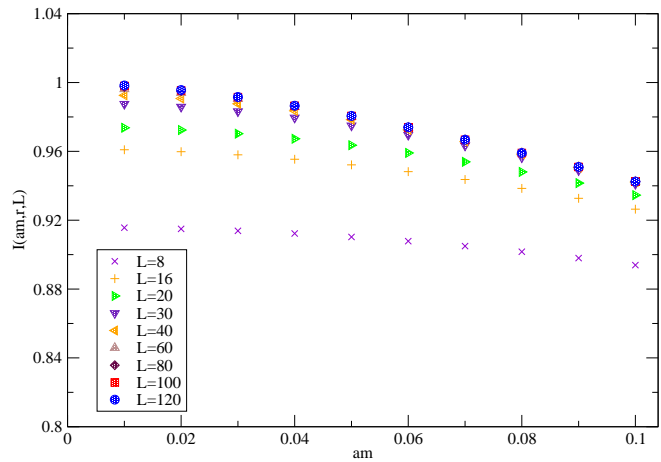}
\caption{The function $I(am,r,L)$ for OStm Wilson fermions 
for the ranges of $am$ 
between 0.01 and 0.1 for different L.}
\label{ostm-Ldep}
\end{figure}

In Fig. \ref{ostm-Ldep} we compare the finite volume dependence of the 
function $I(am,r,L)$ for OStm Wilson fermions at $r=1$ and  
range of $am$ between 0.01 and 0.1. In this range of $am$, convergence is
satisfactory as
volume increases  but the convergence rate becomes slower if the value of $am$ is
smaller.

In conclusion, we have shown that parity violating terms do not arise in the
flavour singlet
axial vector Ward identity up to ${\cal O}(g^2)$ for the OStm
Wilson fermions and the approach to the chiral
limit in comparison with ${\cal O}(a)$, and ${\cal O}(a^2)$
improved Wilson fermions is very satisfactory. 




\begin{thebibliography}{99}

\bibitem{wilson1}
K.~G.~Wilson, $``$Quarks and Strings on a Lattice'', in 
{\em New Phenomena in Subnuclear Physics}, Proceedings of the International 
School of Subnuclear Physics, Erice, 1975, edited by A. Zichichi (Plenum,
New York, 1977). 

\bibitem{karsten-smit}
L.~H.~Karsten and J.~Smit,
Nucl.\ Phys.\  B {\bf 183}, 103 (1981).

\bibitem{kerler1}
W.~Kerler,
Phys.\ Rev.\  D {\bf 23}, 2384 (1981).

\bibitem{frezzotti}
R.~Frezzotti,
Nucl.\ Phys.\ Proc.\ Suppl.\  {\bf 119}, 140 (2003)
[arXiv:hep-lat/0210007].


\bibitem{sint}
S.~Sint,
arXiv:hep-lat/0702008.


\bibitem{shindler}
A.~Shindler,
Phys.\ Rept.\  {\bf 461}, 37 (2008)
[arXiv:0707.4093 [hep-lat]].



\bibitem{osterwalder-seiler}
K.~Osterwalder and E.~Seiler,
Annals Phys.\  {\bf 110}, 440 (1978).

\bibitem{hamber-wu}
H.~W.~Hamber and C.~M.~Wu,
Phys.\ Lett.\  B {\bf 133}, 351 (1983);
H.~W.~Hamber and C.~M.~Wu,
Phys.\ Lett.\  B {\bf 136}, 255 (1984).

\bibitem{wetzel}
W.~Wetzel,
Phys.\ Lett.\  B {\bf 136}, 407 (1984).

\bibitem{eguchi-kawamoto}
T.~Eguchi and N.~Kawamoto,
Nucl.\ Phys.\  B {\bf 237}, 609 (1984).

\bibitem{ETMC09}
P.~Dimopoulos, R.~Frezzotti, C.~Michael, G.~C.~Rossi and C.~Urbach,
  arXiv:0908.0451 [hep-lat].

\bibitem{frezzotti_jhep}
R.~Frezzotti and G.~C.~Rossi,
  JHEP {\bf 0408}, 007 (2004)
  [arXiv:hep-lat/0306014].

\bibitem{seiler-stamatescu}
E.~Seiler and I.~O.~Stamatescu,
Phys.\ Rev.\  D {\bf 25}, 2177 (1982)
[Erratum-ibid.\  D {\bf 26}, 534 (1982)].




\end{thebibliography}
\end{document}